\documentclass[preprint,12pt]{elsarticle}
\let\llncssubparagraph\subparagraph
\let\subparagraph\paragraph
\usepackage[compact]{titlesec}
\let\subparagraph\llncssubparagraph

\usepackage{graphicx}
\usepackage{diagbox}
\usepackage{array}
\usepackage[flushleft]{threeparttable}
\usepackage[numbers]{natbib}
\usepackage{array}
\usepackage{booktabs}
\usepackage{appendix}
\usepackage{makecell}
\usepackage{amssymb}

\usepackage{adjustbox}
\usepackage{multirow}
\usepackage{float}
\usepackage{makecell}
\usepackage{hhline}

\usepackage{xpatch}
\xpatchbibmacro{name:andothers}{%
  \bibstring{andothers}%
}{%
  \bibstring[\emph]{andothers}%
}{}{}
\usepackage{textcomp}
\usepackage[ruled,vlined]{algorithm2e}

\titlespacing{\section}{0pt}{*2}{*1}
\titlespacing{\subsection}{0pt}{*1}{*0}
\titlespacing{\subsubsection}{0pt}{*0.5}{*0}

\journal{Journal of Information Security and Applications}
\begin{document}
\begin{frontmatter}

\setlength{\intextsep}{5pt}%
\setlength{\belowcaptionskip}{-\baselineskip}\addtolength{\belowcaptionskip}{0ex}%

\setlength{\parskip}{0cm}
\setlength{\parindent}{0em}

\title{Lic-Sec: an enhanced AppArmor Docker security profile generator}
%
%
%
\author[1]{Hui Zhu}
\ead{hui.zhu@eit.lth.se}

\author[1]{Christian Gehrmann}
\ead{christian.gehrmann@eit.lth.se}

\address[1]{Department of Electrical and Information Technology, Lund University, Lund, Sweden}

\begin{abstract}
  Along with the rapid development of cloud computing technology, containerization technology has drawn much attention from both industry and academia. In this paper, we perform a comparative measurement analysis of Docker-sec, which is a Linux Security Module proposed in 2018, and a new AppArmor profile generator called Lic-Sec, which combines Docker-sec with a modified version of LiCShield, which is also a Linux Security Module proposed in 2015. Docker-sec and LiCShield can be used to enhance Docker container security based on mandatory access control and allows protection of the container without manually configurations. Lic-Sec brings together their strengths and provides stronger protection. We evaluate the effectiveness and performance of Docker-sec and Lic-Sec by testing them with real-world attacks. We generate an exploit database with 42 exploits effective on Docker containers selected from the latest 400 exploits on Exploit-db. We launch these exploits on containers spawned with Docker-sec and Lic-Sec separately. Our evaluations show that for demanding images, Lic-Sec gives protection for all privilege escalation attacks for which Docker-sec failed to give protection.
\end{abstract}

\end{frontmatter}

\section{Introduction}
Cloud computing is currently a main stream infrastructure technology and different cloud based architectures are used for all kinds of applications and industries. Gartner predicts the cloud computing market to reach \$200 billion in 2019 and it is expected to continue to grow\footnote{https://www.gartner.com/smarterwithgartner/cloud-computing-enters-its-second-decade/}. A key cloud infrastructure technology is the container technology \cite{7036275}. The most popular containerization ecosystems are Docker and Kubernetes which facilitate the management, scaling and deployment of the containerized applications. 

Linux container is a light-weight OS-level virtualization technology. In opposition to virtual machines, containers do not embed their own kernel but run directly on the host kernel \cite{martin2018docker}. The absence of kernel and kernel-sharing property make containers light-weight \cite{guru2019survey}. A problem is that the kernel-sharing property also makes containers not as secure as virtual machines. Virtual machines implement a layer called hypervisor to add an isolation between applications and the host. Containers implement a high-level interface between guest and host, which hides complexity to the user at the cost of becoming more complex itself \cite{williams2018say}. The absence of hypervisor in containers makes them more vulnerable to kernel exploits and attacks on the shared host resources. The authors in \cite{caprolu2019edge} present an overview of several relevant attacks. Thus, container security issues have become a major obstacle for wide adoptions of the containerization technology. To address this issue, Docker implements several Linux kernel security mechanisms such as Capabilities, Seccomp and Mandatory Access Control (MAC). However, even with these default security configurations in place, a Docker container is still vulnerable to many real-world exploits \cite{lin2018measurement}.

One of the most efficient way to prevent such kind of attacks is to apply Linux Security Module (LSM), which is a kernel-level security framework initially targeting Linux \cite{belair2019leveraging}. In this paper, we first studied a recently proposed LSM called Docker-sec \cite{loukidis2018docker}. Docker-sec is a user-friendly security module based on AppArmor to protect Docker containers through their entire lifecycle. It was shown by the authors behind Docker-sec that the proposed LSM is successful in protecting containers from zero-day vulnerabilities with limited performance overhead \cite{loukidis2018docker}. However, so far no-one has evaluated the Docker-sec strength with respect to protection strength against real-world exploits. In this paper, we presents such measurement study. Our evaluations are based on an exploit set extracted from Exploit-DB\footnote{https://www.exploit-db.com}. The exploit set was determined by extracting the exploits in the database which are effective against Docker containers only protected with the default security mechanisms. Then we evaluated the security of Docker-sec by manually executing the exploits on the container platform with Docker-sec enabled. Our results showed that Docker-sec was efficient to attacks exploiting extra network accesses or capabilities in addition to the ones actually required for proper function. In other conditions when the exploited capabilities and networks are truly required by the application running in the container, Docker-sec was not efficient. To address this issue, we then introduced the new AppArmor profile generator tool, Lic-sec. The tool was designed by combining Docker-sec, with a different, earlier proposed LSM tool called LiCShield \cite{mattetti2015securing}. We have adapted and extending LicShield to fit the Docker-sec profile generation principles. The effectiveness of Lic-Sec was evaluated by using the very same exploit set as we used for the Docker-sec evaluation. Our measurements shows that Lic-sec manages to give much higher protection level. In particular, it prevents all privilege escalation attacks which Docker-sec failed to mitigate.

In summary, we make the following contributions:
\begin{description}
  \item[$\bullet$]We have evaluated how Docker-sec  performs against real-world attacks.
  \item[$\bullet$]We have constructed a new AppArmor profile generator tool that combined Docker-sec with a modified version of an older AppArmor profile generator, LiCShield, which we call Lic-Sec.
   \item[$\bullet$]We show that the new tools performs better than the pure Docker-sec tools. In particular we show that it gives protection against all privilege escalation attacks which Docker-sec could not defend against.
   \item[$\bullet$]We present performance figures for the new combined profile generation tool.
\end{description}

The rest of this paper is organized as follows. In Section 2, we give a background to the Docker Architecture, Mandatory Access Control, Docker-sec and LiCShield. In Section 3, we formulate the main research problem, i.e, the evaluation goal of Docker-sec and the design goal of the AppArmor profile generator. In Section 4, we introduce the methodology used in the evaluation and test setup and we describe how the exploit database was generated. In Section 5, the main Docker-sec security evaluation results are presented and a detailed analysis of the results is given. In Section 6, we describe the design details for Lic-Sec. In Section 7, the results for security and performance evaluations for Lic-Sec are introduced as well as an analysis for the enhancements of Lic-Sec compared to pure Docker-sec. In Section 8, we present and discuss Docker security related work. In Section 9, we conclude the evaluation study and identify future work.

\section{Background}
In this section we give a brief background to the Docker technology and architecture. In addition, we give an overview of key Linux and container security concept with a focus on the security technologies behind Docker-sec and LiCShield.
\subsection{Docker Architecture and Components}
\begin{figure}[h]
\centering
    \includegraphics[width=0.7\textwidth]{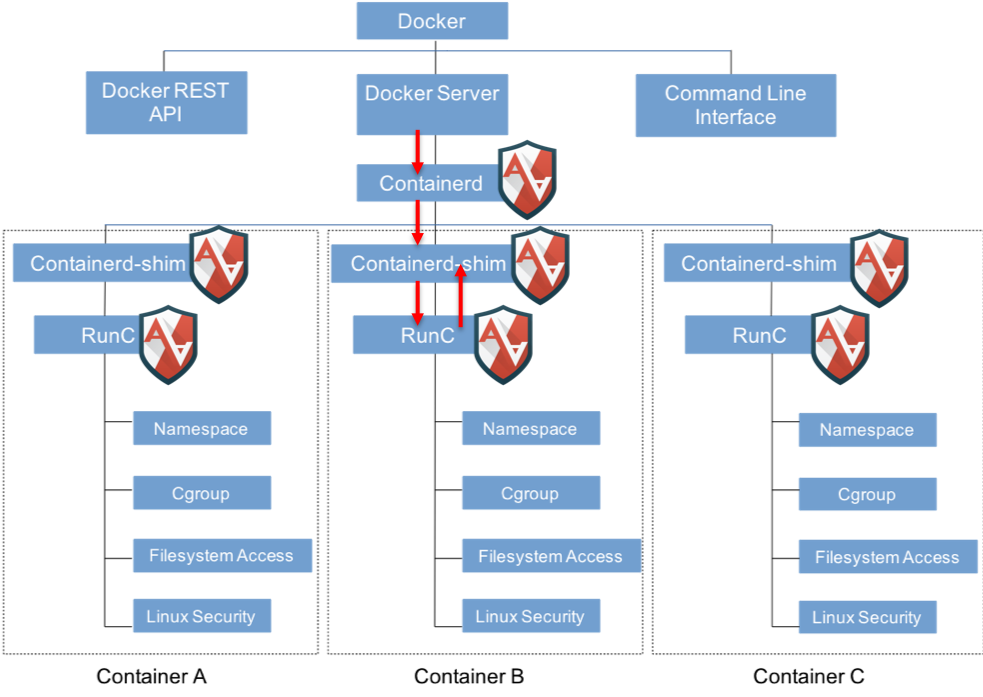}
    \caption{Docker Architecture and Critical Components}
    \label{fig:DockerArch}
\end{figure}

Docker uses a client-server architecture which consists of a server, a REST API and a command line interface (CLI) client as shown in Figure \ref{fig:DockerArch}. It can be further broken into four major components: the Docker server, the Containerd, the Containerd-shm and the RunC\footnote{https://docs.docker.com/get-started/overview/}.

The Docker server is a long-running program which is also called the Docker daemon. It takes responsibility for the control, creation and management of Docker objects such as containers, images, networks and volumns. The Docker daemon implements the Containerd to manage the life-cycle of containers. The Containerd uses the RunC to run containers based on Open Container Initiative (OCI) specifications. The RunC is a command-line tool for spawning and running containers. It creates a container using Namespaces, cgroups, filesystem access controls and Linux security capabilities. After the container actually runs, the RunC exits and hands the control over to the Containerd-shim, which sits between the Containerd and the RunC. The shields in Figure \ref{fig:DockerArch} indicates the components in the architecture guarded by Docker-sec. 

\subsection{Mandatory Access Control (MAC)}
In Mandatory Access Control (MAC) systems, mandatory policies govern access on the basis of classification of subjects and objects in the system. Each subject and object in the system are assigned a security level. The security level associated with an object reflects the sensitivity of the information contained in the object. The security level associated with a subject, also called clearance, reflects the subject's trustworthiness not to disclose sensitive information to subjects not cleared to see it \cite{sandhu1994access}. There are two kinds of tools for MAC in Linux: AppArmor and SELinux. SELinux is label-based and uses a type enforcement model. In SELinux, types are associated to applications and resources to enforce access rules \cite{Mayer2006}. Since Docker-sec evaluated in this paper is based on AppArmor, we will not discuss SELinux further. Different from SELinux which is based on label, AppArmor is based on paths. More details of AppArmor and its usage in Docker container are given in Section 2.3 below.

\subsection{AppArmor} 
AppArmor \cite{dockersecurityapparmor} is integrated in Ubuntu/Debian distributions and it is used by Docker. Users can define a specific AppArmor profile for a single application, restricting what it can do and cannot do based on paths. In this way, even if the attacker has succeeded in exploiting vulnerabilities in an application, the behavior of the compromised application is still restricted by AppArmor.

When a container is started by Docker, a default AppArmor profile named 'docker-default' will be generated by Docker binary and loaded into the kernel for the container runtime. There is also an AppArmor profile for the Docker daemon but it is currently not installed with Docker\footnote{https://docs.docker.com/engine/security/apparmor/}. Generally, the default pre-defined AppArmor profile for container is moderately protective to provide wide application compatibility. The default profile for instance forbids mount operations. It also denies access to important file systems on the host \cite{bui2015analysis}. However, the default Docker AppArmor profile does not restrict the network and capability rules. This brings high risk of the host suffered from privilege escalation attacks through a compromised container. Besides the pre-defined profile of Docker, administrators can also define and implement their own profiles which provides more strict restrictions. This requires advanced configuration skills by the person responsible for deploying the Docker application.

\subsection{Docker-sec}
\label{tab:Docker-sec}
The authors in \cite{loukidis2018docker} propose a novel security mechanism for Docker based on AppArmor which is called Docker-sec. Docker-sec adds an additional security layer on top of Docker's security defaults by automatically creating per-container AppArmor profiles.

The AppArmor policy improved by Docker-sec has three major advantages compared to the default AppArmor policy loaded and enforced by Docker: 1) It protects the container through its whole life-cycle by generating secure profiles for all the critical Docker components as shown in Fig.\ref{fig:DockerArch}. The default AppArmor profile only protects the container after the initialization by the RunC. 2) It generates per-container AppArmor profiles rather than the common secure profile compatible to all containers. 3) The AppArmor profile for each container is dynamic and can be adapted to any behavior change of the container by collecting the behavior of the container during training period and then determining the privileges that are truly necessary for the container to run properly.

\subsection{LiCShield}
LiCShield framework was presented in \cite{mattetti2015securing} for securing Docker containers and their workloads by automatically generating AppArmor rules for both the host and the Docker container. In brief, LiCShield traces all kernel operations by the trace tool called SystemTap\footnote{https://sourceware.org/systemtap/} while the Docker daemon is performing the build and run operations. LiCShield translates all the traces to AppArmor rules and constructs two different AppArmor profiles: one targeted to the operations performed inside the container, the other one to those performed on the host. 

LiCShield is similar to Docker-sec in design principles since both of them provide automatic AppArmor profile construction but with the following differences as listed in Table \ref{tab:Comparison}. First, we explain the differences that highlight the strengths of LiCShield: 1) LiCShield uses SystemTap as the tracing tool. This tool is more flexible than Auditd used by Docker-sec as it allows customizing the tracing script according to user's needs. Users can monitor and investigate a variety of kernel functions, system calls\footnote{https://sourceware.org/systemtap/SystemTap\_Beginners\_Guide/scripts.html} and process-related activities\footnote{https://sourceware.org/systemtap//man/tapset::kprocess.3stap.html}, etc. This feature enhances the extensibility of LiCShield; 2) LiCShield generates important rules which Docker-sec cannot generate, including pivot root rule, access rule, mount rule, link rule and execution rule. These rules could be used as the supplementary to the rules generated by Docker-sec. 3) LiCShield automatically constructs the profile for the Docker daemon based on its executions, which confines the privilege to the bare minimum needed for the proper function. While Docker-sec adopts a pre-defined AppArmor profile for the Docker daemon, which is moderately protective. 

Besides the strengths, LiCShield has several weaknesses compared to Docker-sec since LiCShield is incompatible with the latest Docker releases\footnote{https://github.com/docker/docker-ce/releases/tag/v19.03.8}: 1) LiCShield does not generate any profile for the RunC;  2) LiCShield does not give good runtime container protection. This is caused by the principles behind how  LiCShield enforces the container protection profile. LiCShield exploits the pivot root rule in the host profile to enforce the container protection. In earlier Docker versions, the Docker daemon calls the pivot root right before starting the container processes. Hence, for older Docker versions, this it is the right point to switch from the host profile to the container profile. But in current Docker versions, the RunC takes over this task from the Docker Daemon. Therefore, the pivot root rule cannot be used anymore for this purpose.
\begin{table}[h]
\centering
    \caption{Comparison between Docker-sec and LiCShield }
    \label{tab:Comparison}
    \resizebox{.98\textwidth}{!}{
    \begin{tabular}{|c|c|c|c|c|c|c|c|}
    \hline
        LSM & year & MAC & tracing tool & generated rules & generated profiles & pre-defined profiles & effective protective range  \\ \hhline{|=|=|=|=|=|=|=|=|}
        Docker-sec & 2018 & AppArmor & Auditd & {\begin{tabular}{@{}c@{}}capability rule \\ network rule\end{tabular} } & container runtime & {\begin{tabular}{@{}c@{}}RunC \\ Docker Daemon\end{tabular} } & {\begin{tabular}{@{}c@{}c@{}}container \\ RunC \\ Docker Daemon\end{tabular} } \\\hline
        LiCShield & 2015 & AppArmor & SystemTap & {\begin{tabular}{@{}c@{}c@{}c@{}c@{}}pivot root rule \\ access rule \\ mount rule \\ link rule \\ execution rule \end{tabular} } & {\begin{tabular}{@{}c@{}}container runtime \\ Docker Daemon\end{tabular} } & / & Docker Daemon \\\hline          
    \end{tabular}
   }
\end{table}

\section{Problem description}
We believe AppArmore automatic profile generation has great potential in enhancing the security of containers. It is a major advantage if the profiles can be automatically generated in order to avoid forcing the end-user to make cumbersome manual configuration. However, despite this fact, it is an open research question to measure the security strength of AppArmor profiles. There is also a need to design enhanced automatic profile generation tools providing good protection.

The aim of our research is to provide a quantitative evaluation of the effectiveness of automatically generated AppArmor profiles in protecting containers against different types of exploits, in order to identify the strengths and weaknesses of the profiles and find an enhanced profile generator to avoid the weaknesses. To accomplish this goal, we break the problem down into the following three sub-problems: 1) find a methodology and test framework for evaluating the strength of automatically generated AppArmor profiles for Docker containers; 2) use the methodology and test framework to evaluate the strength of Docker-sec and identify potential weaknesses; 3) find an enhanced profile generator offering higher security than Docker-sec.

\section{Experimental setup}
Next, we describe the experimental set-up we have used in our evaluation. First we discuss how we classify attacks, then we describe how we have collected the exploit used in the tests and finally we explain how the actual tests were executed.
\subsection{Attack Classification}
We have chosen to use a similar two-dimensional method for classification as suggested by Lin et al. in \cite{lin2018measurement} but with the following differences:
1) We choose the targeting objects of the attacks as the first dimension which is slightly different from the influence range used in \cite{lin2018measurement} . We consider this way more intuitive and straight-forward since the targeting objects could be obtained directly from the official descriptions of vulnerabilities in National Vulnerability Database (NVD).
2) We also choose the impacts of the attacks as the second dimension but with different approaches, i.e, we determine the impacts of the attacks based on the vulnerability type for each attack provided by NVD as shown in Table \ref{tab:classification}.

First, we classify the exploits based on the targeting objects, i.e., Web Application, Server, Database and Kernel. These are obtained from the "Known Affected Software Configurations" of vulnerability provided by NVD. Second, we classify the exploits into five categories based on the impacts of attacks. In order to summarize the impacts of all the attacks, we manually analyze the vulnerability type for each attack provided by NVD. The vulnerability type is defined by CWE (Common Weakness Enumeration)\footnote{https://cwe.mitre.org/data/index.html}. For each type, CWE defines its negative technical impact when the exploit is successfully used. We classify similar impacts into the same category. Then we obtain five categories: Bypass, Gain Privilege, Denial of Service (DoS), Gain Information and Execute Code, which are listed in Table \ref{tab:classification}. Compared to the Lin et al. work, we introduce one more category called "bypass", which is explained in details here.  We here also further discuss privilege escalation attacks, which is the most important security class with respect to container security.  
\begin{table}[h]
\centering
    \caption{Classification of different attack types}
    \label{tab:classification}
    \resizebox{.5\textwidth}{!}{
    \begin{tabular}{|l|l|}
    \hline
       Impact & Category \\ \hhline{|=|=|}
        Bypass Protection Mechanism & Bypass\\ \hline
        Gain Privileges or Assume Identity & Gain Privilege \\\cline{1-1}
        Modify Application Data & \\\cline{1-1}
        Modify Files or Directories & \\  \cline{1-1}
        Modify Memory & \\  \hline   
        DoS: CPU Resource Consumption & {\begin{tabular}{@{}c@{}}Denial of Service \\ (DoS)\end{tabular} }\\\cline{1-1}
        DoS: Memory Resource Consumption  & \\\cline{1-1} 
        DoS: Resource Consumption (Other) & \\\cline{1-1} 
        DoS: Crash, Exit, or Restart & \\\cline{1-1} 
        DoS: Instability & \\ \hline          
        Read Application Data & Gain Information\\\cline{1-1} 
        Read Memory & \\\cline{1-1} 
        Read Files or Directories & \\  \hline          
        Execute Unauthorized Code or Commands & Execute Code\\  \hline
    \end{tabular}
    }
\end{table}
\begin{itemize}
    \item \textbf{Bypass: }By bypassing a protection mechanism, an attacker gain unauthorized access to a system. There are two general ways for attackers to bypass a protection mechanism; either gaining unauthorized access to a system by exploiting insecure interaction between components or by compromising security functionality by risky resource management. The first way is commonly used in injection attacks and Cross-Site Request Forgery (CSRF) attacks. The second way is commonly used in DoS type of attacks. 
    \item \textbf{Gain Privilege: } Kernel vulnerabilities are used by attackers to gain unauthorized privileges of the container and the host. Depending on the effective range of the gained privilege, there are two kinds of privilege escalation attacks:
   \begin{itemize}
        \item \textbf{Inside container:}  Docker provides 14 capabilities for the root user in containers instead of 38 capabilities for the real root user in the host. However, an attacker can bypass the CPU protection mechanisms and exploit kernel vulnerabilities to get the real root privilege, i.e., to obtain the full 38 capabilities. When using an exploit in this category, the attacker is still restricted inside the container and has no access to the underlying host and other containers running on the same host.
        \item \textbf{Container escape:} In a successful container escape attack, the attacker not only gets the real root inside the container, but also breaks the isolation between containers and the underlying host. After a successful attack, it is possible for the attacker to modify critical files or memory of the host and further influence or change other containers through the host.  For example, CVE-2019-5736\footnote{https://nvd.nist.gov/vuln/detail/CVE-2019-5736}, a vulnerability of RunC, allows an attacker to overwrite the host RunC binary and consequently obtain host root access. Another example is a Namespaces switching attack which we describe in more details in the analysis in Section \ref{PrivEscAttack}.
    \end{itemize}
    \item \textbf{Denial of Service (DoS): } A DoS attack results in resource exhaustion or disability. An attacker can utilize a DoS attack to slow down or crash the software and deny service to legitimate users.
    \item \textbf{Gain Information: }Attackers can read the critical application data, files, directories and memory to gain sensitive information such as credentials, passwords and keys. 
    \item \textbf{Execute Code: }A remote attacker can execute arbitrary codes on a victim's computer, generally targeting on Web Applications and servers. 
\end{itemize}

o\subsection{Exploit Database Collection}
We generated our exploit database collection using similar principle as presented in \cite{lin2018measurement} but modified to suit the Docker-sec evaluation target. In \cite{lin2018measurement}, the authors collected the latest 100 exploits of each category from Exploit-db\footnote{https://www.exploit-db.com} and filtered out the exploits which would probably fail on the container platform such as the ones attacking the graphic user interface related program. Finally, they got 223 exploits out of 400 exploits which might be effective on container platform. In our evaluation, the purpose is to obtain the exploits which are actually effective on container platform with Docker default security mechanism. To achieve this, we processed the exploit database through three steps. First, we generated the universe set of exploits from Exploit-db. Second, we did a first-round filter to filter out the exploits which are not commonly deployed in containers. Third, we applied a second-round filter to filter out the exploits which failed to launch the attacks on containers with just default security mechanisms enabled. 
\begin{itemize}
    \item \textbf{Generation of universe set of exploits: }We used the same method as \cite{lin2018measurement} to generate the original 400 exploits collection from Exploit-db. Exploit-db is the oldest and most comprehensive database of real-world exploit code for several platforms such as Windows, Linux, etc. Exploit-db divides the exploits into four categories: Web Application, remote, local \& privilege escalation, and denial of service. Based on the four categories, we collected the latest 100 exploits of each category targeting on Linux platform before December 31, 2019. Among the 400 exploits, 126 exploits were published in 2019. 190 exploits were published in 2018. 75 exploits were published in 2017 and 2016. Only 9 exploits were published before 2016.
    \item \textbf{First-round filter: }The purpose with this step was to obtain the exploits which might be effective on container platforms. We used theoretical investigations mainly focusing on exploits of user space programs. We studied the text description of each exploit manually to confirm the affected user space program. Then we investigated the application based on the following criteria: 1) The exploits targeting graphic user interface (GUI) applications were filtered out since containers are generally used for deploying back-end services such as web server and database server; 2) The exploits targeting back-end applications, which are not generally used in containers, were filtered out since containers are designed for hosting single service in a single process. As recommended by Docker developers \cite{combe2016docker}, it is not suitable to deploy an application running as a platform which integrating several applications. For example, EDB-ID 46785\footnote{https://www.exploit-db.com/exploits/46785} targeting Ruby On Rails were filtered out. This is a Web Applicationlication framework that includes everything needed to create database-backed Web Applicationlications.
    \item \textbf{Second-round filter: }The purpose with the second step was to obtain the exploits that bypass all the default Docker protection mechanisms. These exploits are the most severe threats against the container and by testing them, we get a good view of the strengths and weaknesses of Docker-sec. First, we analyzed the exploit code and text description of CVE to figure out the attack principle and necessary conditions to carry out the attacks. Then we implemented the exploit code on a Docker container to see if the exploit was actually effective with the default Docker security configurations. We used the same method as in \cite{lin2018measurement}  for the exploits targeting user space programs, i.e we deployed the vulnerable programs inside the container and ran the exploits outside the container. For the exploits which target the Linux kernel, we deployed the vulnerable kernel on the host and executed the exploits inside the container.
\end{itemize}
After the two-round filter of the universe set of exploits, we obtained the final exploit database collection for our evaluation. For each exploit, we collected the EDB ID (Exploit Database IDentifier), CVE (Common Vulnerabilities and Exposures IDentifier) ID, publishing date, CVE type, exploit code to launch the attack, targeting object and text description of CVE. Then we utilized the classification method described in Section 3.1 to classify the exploit database. The number of exploits in the final database according to the classification is listed in Table \ref{tab:database}.

\begin{table}[h]
\centering
\resizebox{.65\textwidth}{!}{
\begin{threeparttable}
    \caption{Exploit Database Collection}
    \label{tab:database}
    \small
    \begin{tabular}{|c|c|c|c|c|c|}
    \hline
        \backslashbox[8.7em]{Categories}{Objects}  & Web Application & Server & Database & Kernel & Total \\ \hhline{|=|=|=|=|=|=|}
        Bypass                                                      & 6 & 3 & $-$  & $-$ & 9    \\ \hline
        \begin{tabular}{@{}c@{}}Gain Privilege \\ (Inside Container)\end{tabular}     & $-$ & $-$ & 2 & 4  & 6    \\ \hline
        \begin{tabular}{@{}c@{}}Gain Privilege \\ (Container Escape)\end{tabular}     & $-$ & $-$ & $-$ & 4 & 4    \\ \hline
        DoS                                                         &  3 & 1  &  $-$ & 1 & 5     \\ \hline
        \begin{tabular}{@{}c@{}}Gain \\ Information \end{tabular}   & 2 & 2 & 1  & $-$ & 5    \\ \hline
        Execute Code                                                & 15 & 2 & $-$ & $-$ & 17    \\ \hline
        Total                                                       & 22 & 8 & 3 & 9 & $42^*$    \\ \hline
    \end{tabular}
\begin{tablenotes}
\small
\item *: Total number is less than the sum of all rows in the column. This is because some exploits cause more than one consequences.
\end{tablenotes}  
\end{threeparttable}
  }
\end{table}

\subsection{Test Setup}
We used Docker 19.03.1-ce for the evaluations. This version was released on 25th July 2019 and supports Linux kernel security mechanism including Capability, Seccomp and MAC. Furthermore, on the host we ran the Linux distribution Ubuntu 14.04 LTS with kernel version 4.4.0-51-generic. This kernel version was chosen to guarantee that the host kernel is vulnerable to all the kernel vulnerabilities in the selected exploit collection. This is obviously a requirement in order to do the security evaluation based on existing exploits.

Docker-sec can generate the container profile through two mechanisms: static analysis and dynamic monitoring. In the static analysis mode, the profile is generated based on information about the container and the type of access rights it needs, which is either provided by the users as the command line arguments or generated by Docker. However, customizing these configurations is very time-consuming and if the user does not customize \textsl{docker run}, the resulting profile will be the same as the default AppArmor profile provided by Docker. 

Therefore, in our test, we directly enabled the dynamic monitoring for Docker-sec and we saved the runtime profile for each tested application for further analysis. We enabled dynamic monitoring by executing \textsl{docker-sec train-start} and \textsl{docker-sec train-stop} command. During this two commands we performed all the required application functionalities. Then Docker-sec replaced capabilities and network accesses in the initial profile generated by static analysis, with the necessary capabilities and network accesses collected during the training period.

\section{Evaluation of Docker-sec}
Now, we present the exploit test results obtained in the study. We start by discussing the overall results and then we put special attention to the privilege escalation attacks, which we treat in a separate sub section. 
\subsection{Test Result Analysis}
Based on the targeting objects, we divide the exploits into two categories: user space program targeted exploits and kernel targeted exploits. User space targets include Web Applicationlications, server and database while the only kernel target considered is the  Linux kernel. The results are displayed in Table~\ref{tab:result1}.
\begin{itemize}
    \item \textbf{User space program targeted exploits: } Some user space targeted exploits, which require Linux capabilities and network accesses to launch, can be defended by Docker-sec. Since we 'train' the runtime profile by Docker-sec and discard the unnecessary capabilities and network accesses, not all required conditions to launch some of these attacks are fulfilled. For example, EDB-40968\footnote{https://www.exploit-db.com/exploits/40968} exploits CVE-2016-10033 to launch a remote code execution against PHPMailer. With the default runtime profile, the attack was successful. However, after training the container with Docker-sec, the capabilities were limited to 'setuid' and 'setgid' only, and the network accesses were constrained to 'netlink raw', 'inet stream', 'inet dgram', 'inet6 stream' and 'inet6 dgram'. The attack failed under these restrictions due to the lack of network accesses: network unix dgram and network unix stream.

    The rest of the application-targeted attacks succeeded even with the Docker-sec generated AppArmor profile. The main reason is that these exploits do not directly depend on Linux capabilities or network accesses. Instead, these attacks generally exploits some common Web Application vulnerabilities such as injection, cross-site request forgery and unrestricted upload of file with dangerous type. It is worth to notice though, that even if this is the case,  Docker-sec is still effective in the sense that attack is isolated to the container and will not affect any program running outside of it. 
    \item \textbf{Kernel targeted exploits:} When it comes to kernel-targeted exploits, Docker-sec is much more effective as it restricts container capabilities. In particular, in the cases when the users run the Docker container on a host with kernel version 4.4.0. As Docker-sec reduces the capabilities to those really needed, several attacks can be prevented. However, clearly, if some rules are truly required during container runtime, then Docker-sec cannot discard these capabilities opening up for kernel-targeted exploits. In order to evaluate the defense performance in real-life application scenarios, we test the top 20 official Docker images pulled from Docker hub\footnote{https://hub.docker.com/search?q=\&type=image} with Docker-sec and we found that only two images, 'Docker in Docker' and 'IBM DB2' are vulnerable to the majority of the attacks. For the remaining tested 18 images, the Docker-sec generated profile gives the expected protection as also can be seen in Table \ref{tab:result1}. We analyze the failed cases in further details in Section 5.2 below.
\end{itemize}

\subsection{Analysis of Privilege Escalation Attacks}
\label{PrivEscAttack}
Once a kernel-targeted exploit is launched successfully inside a container, the attacker will get root privilege and will be able to break the isolation by escaping from the container to the host. This is the far most severe type of attack and in this section we dig deeper into the analysis and test results with respect to privilege escalation. Our focus is to analyze the efficiency of Docker-sec in restricting the attack ranges and minimizing the impacts. As we described in Section 4.1, there are two types of privilege escalation attacks. Below, we discuss the test results with respect to these two types separately.
\begin{itemize}
    \item \textbf{Inside container: } We tested 4 exploits. Among these, 3 of them (CVE-2017-7308\footnote{https://www.exploit-db.com/exploits/41994},  CVE-2017-6074\footnote{https://www.exploit-db.com/exploits/41458} and CVE-2017-1000112\footnote{https://www.exploit-db.com/exploits/43418} require capability SYS\_ADMIN to launch and 1 (CVE-2016-9793\footnote{https://www.exploit-db.com/exploits/41995}) requires capability NET\_ADMIN. If none of these capabilities are given to the container, all of these attacks fail. This is the reason Docker-sec protects the 18 tested images from the 4 exploits. During the images' runtime, Docker-sec discards the NET\_ADMIN and SYS\_ADMIN since they are not necessary for the images to run properly. However, for the 'Docker in Docker' and 'IBM DB2' images, Docker-sec adds these two capabilities into the runtime profile after the training period and consequently the privilege escalation attacks succeed.

    Docker-sec is not any more effective after privilege escalation inside the container. In these attacks, once the real root privilege has been obtained inside the container, a root shell with a new PID is spawned. This process with real root privilege is not enforced by Docker-sec runtime profile and Docker-sec has no protection effect anymore. However, it should be noted that even if the capability restriction and MAC under these attack circumstance not are effective anymore, the Namespaces isolation and Docker Seccomp profile protections are still working. This means that the default read-only restriction for file access still applies despite the attacker root access. Therefore, the attack only affects resources inside the container and the host remains unaffected.
    
    \item \textbf{Container escape: } Typically, the attacker switches the Namespaces to the one of the host or hosts to launch the container escape attack. We tested 4 exploits based on this mechanism. As there is no 'ready-made' exploit code to launch these attacks, we used a well proven technology\footnote{https://www.cyberark.com/threat-research-blog/the-route-to-root-container-escape-using-kernel-exploitation/} to test this attack procedure. The root kernel exploits which we based our tests on were  CVE-2017-1000112, CVE-2016-9793, CVE-2017-7308 and CVE-2017-6074. Among these exploits, three exploits need capability SYS\_ADMIN to launch and one exploit needs capability CAP\_ADMIN to launch. The attack procedure we used is as follows. First, the attacker bypasses the Kernel Address-space Layout Randomization (KASLR) mechanism to get the kernel base address. Since the offsets of kernel symbols are constant to the kernel base address, it is easy to obtain the kernel symbol addresses through /proc/kallsyms which will be used in the container escape attack. Next, the attacker calls the kernel function find\_task\_by\_vpid to obtain the task\_struct of virtual process 1, which is the first process inside the container. Then the attacker further calls the switch\_task\_Namespaces() function to change attackers' process's Namespaces to those of the host's. The input parameters for switch\_task\_namespaces() function are task\_struct of process 1, which is obtained in the previous step, and the address of init\_nsproxy structure which contains the Namespaces of the initial process of the host. Finally, the attacker calls the setns syscall from the kernel exploit to perform the required Namespaces changes and then escape to the host.
    The defending principle of the container escape attacks is the same as the privilege escalation inside containers as we previously described. Docker-sec discards those two capabilities for the 19 images runtime. The cause of failure for image 'Docker in Docker' and 'IBM DB2'is also the same. Docker-sec considers these two capabilities as necessary and adds them into the runtime profile. 

\end{itemize}

\begin{table}[h]
\centering
\resizebox{\textwidth}{!}{
  \begin{threeparttable}
    \caption{Docker-sec Security Evaluation Result Overview}
    \label{tab:result1}
    \small
    \begin{tabular}{|c|c|c|c|c|c|}
    \hline
        Exploits &  \multicolumn{3}{c|}{Userspace program targeted} & \multicolumn{2}{c|}{Kernel targeted} \\ \hhline{|=|=|=|=|=|=|}
        \backslashbox[8.7em]{Categories}{Objects}  &\begin{tabular}{@{}c@{}}Web Application \\ ($Doc/Sec^1$)\end{tabular} & \begin{tabular}{@{}c@{}}Server \\ ($Doc/Sec^1$)\end{tabular} & \begin{tabular}{@{}c@{}}Database \\ ($Doc/Sec^1$)\end{tabular} & \begin{tabular}{@{}c@{}}Other images \\ ($Doc/Sec^1$)\end{tabular} & \makecell{Docker in Docker\\IBM DB2\\ ($Doc/Sec^1$)} \\ \hline
        Bypass                                                      & 6/6 & 3/3 & $-$  &$-$ &$-$ \\ \hline
        \begin{tabular}{@{}c@{}}Gain Privilege \\ (Inside Container)\end{tabular}     & $-$ & $-$ & 2/2 & 4/0 & 4/4    \\ \hline
        \begin{tabular}{@{}c@{}}Gain Privilege \\ (Container Escape)\end{tabular}     & $-$ & $-$ & $-$ & 4/0 & 4/4\\ \hline
        DoS & 3/3  & 1/1 & $-$  & 1/0 & 1/1 \\ \hline
        \begin{tabular}{@{}c@{}}Gain \\ Information \end{tabular}   & 2/2 & 2/2 & 1/1  & $-$ &$-$ \\ \hline
        Execute Code                                                & 15/14 & 2/2 & $-$ & $-$ &$-$\\ \hline
        Total                                                       & 22/21& 8/8 & 3/3 & 9/0 & 9/9 \\ \hline
    \end{tabular}
    \begin{tablenotes}
      \small
      \item 1: 'Doc' denotes the number of exploits executed successfully on containers launched with Docker, 'Sec' denotes the number of exploits executed successfully on containers launched with Docker-sec
     
    \end{tablenotes}
\end{threeparttable} }
\end{table}

\section{Lic-Sec - combining Docker-sec with LiCShield}
Now we proceed with describing our enhancements to Docker-sec with the purpose of getting an overall stronger profile generator. We call our new tool, Lic-Sec, an AppArmor profile generator which combine Docker-sec and LiCShield. First we discuss the Lic-Sec design motivations. Then, we show the modifications we have done to LiCShield as well as their purposes. Next, we describe how Lic-Sec works and the main differences compared to Docker-sec. In the end, we explain how to use the tool in different user scenarios.
\subsection{The design of Lic-Sec}
The design of Lic-Sec is motivated by the evaluation results of Docker-sec. As concluded in Section 5.1 and 5.2, the Docker-sec limitation with respect to only restricting capabilities or networks imply that we cannot defend against some of the attacks. Based on the analysis of the differences between LiCShield and Docker-sec discussed in Section 2.5, we notice that if one try to combine LiCShield with Docker-sec, we will be able to give a higher level of protection and potential addressing the shortcomings identified with respect to Docker-sec defense strength.  Especially,  LiCShield has the potential to collect more system information and generate additional protection rules (compare to Docker-sec) such as pivot root rules, file access rules, mount rules, link rules and execution rules. Those rules generated for the container could be combined with the capability and network rules generated by Docker-sec to construct a profile with more strict restrictions, broader container run-time analysis, giving and overall more flexible tool. 

In summary, the design goals for the combined tool are: 1) automatic construction of AppArmor profiles for per-container and all the Docker components; 2) more strict confinement of the privileges inside the container; 3) support for extensibility of functionality

\subsection{The modifications of LiCShield}
Based on the design goal of Lic-Sec, we made the following modifications of LiCShield. 
\begin{enumerate}
    \item We modified LiCShield’s approach of enforcing container profile in order to enforce per-container profile with more rules. We changed the way AppArmor profiles are enforced. Instead of using the LicShield approach where a profile is enforced with a pivot root rule on the host, we used the LicShield profile generator but applied them using the Docker sec enforcement principle.
    \item We modified LiCShield’s SystemTap script in order to trace all processes triggered by different Docker components. First, we add one more field to the trace log structure called 'executable process name' in order to identify processes with empty executable path such as 'runc:[2:INIT]'. Second, we enable the trace for processes also for processes which are not in the target tree.
    \item We modified LiCShield’s rules generator from two aspects: first, we generate deny rules for prompting shells inside container in order to defend privilege escalation attacks. If no command such as ’/bin/bash’,’/bin/sh’ and ’/bin/dash’ is tracked during the tracing period, operations of ’r’ (read), ’w’ (write), ’m’ (memory map as executable), ’k’ (file locking), ’l’ (creation hard links) and ’x’ (execute) to those commands are all denied by adding the deny rules to the container profile. Second, we change the way of classifying generated rules to container profiles and host profiles in order to identify profiles more accurately.
    \item We modified Docker-sec’s training process and rules generator in order to automate the generation of mount rules. We add auditing for mount operations when training process starts in addition to the Docker-sec network and capabilities. We also add a function for generating mount rules based on the Auditd log.
\end{enumerate}

\subsection{The mechanisms of Lic-Sec}
\begin{figure}[h]
\centering
    \includegraphics[width=0.98\textwidth]{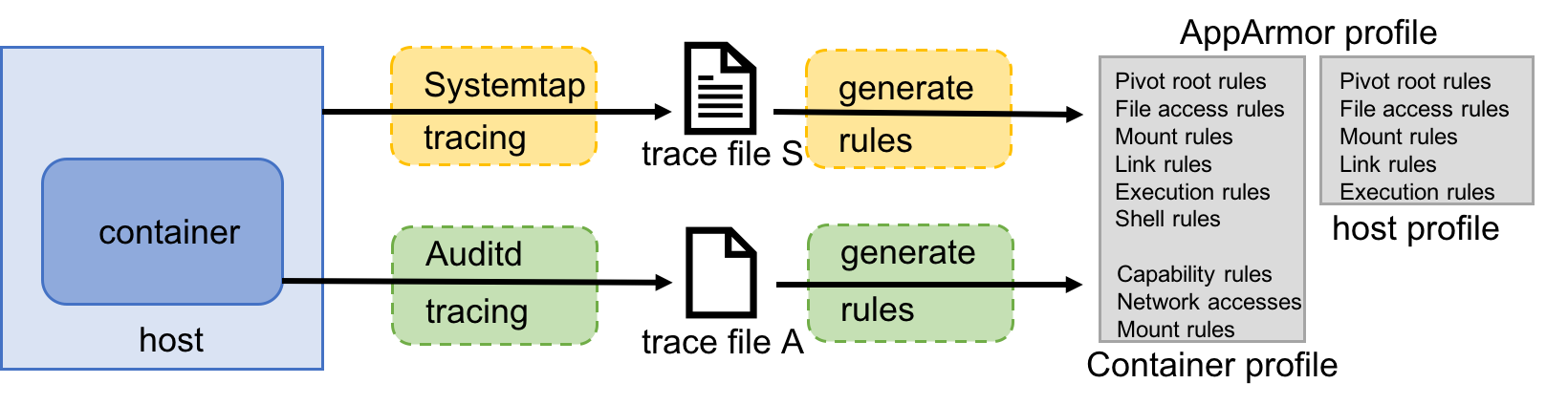}
    \caption{Lic-Sec Approach Overview}
    \label{fig:approach}
\end{figure}

Here we list the main mechanisms of Lic-Sec which include tracing and profile generation. The whole process is displayed in figure \ref{fig:approach}.

1) tracing: Tracing is the procedure of the new profile generator to collect necessary information as the basis to generate rules. SystemTap and Auditd are leveraged as the tracing tools. When the tracing starts, SystemTap and Auditd work in parallel: the former one collects kernel operations performed by all the Docker components based on a pre-defined trace script and the latter one collects mount operations, capability operations and network operations performed by the running container. The structure of SystemTap trace can be customized and flexible while the structure of Auditd trace is defined by Auditd and fixed. We design the structure for SystemTap trace based on LiCShield. We add one more field called executable process name (marked with a star) in addition to LiCShield. The new structure has the following fields:\\

\textlangle{}probe point name\textrangle{}  \textlangle{}control group path\textrangle{}  \textlangle{}executable process name\textrangle{} \textlangle{}executable path\textrangle{}  \textlangle{}resources path\textrangle{}  \textlangle{}mount namespace root\textrangle{}. 
\begin{itemize}
    \item \textbf{probe point name:} the name of the kernel function probed, which identifies the required privileged operation. For example, when 'security\_sb\_pivotroot' function is probed, the pivot\_root privilege is required. Lic-Sec identifies the type of the rule depending on this field. 
    \item \textbf{control group path:} identifies the source of the request, either from the container or from the host. This is used for distinguish the operations between inside the container and the host and classify the generated rules to container profile and host profile. 
    \item \textbf{executable process name*:} the name of the process which performs the operation. This field is introduced since we could not identify the process from the executable path for some processes such as the RunC:INIT. 
    \item \textbf{executable path:} the executable path of the process which performs the operation, acting as the subject of the privilege operation.
    \item \textbf{resources path:} the resource involved in the operation, acting as the object of the privilege operation. 
    \item \textbf{mount namespace root:} identifies the involved resource's mount point, which could also be used for identifying where the request comes from. Since if the operation is performed inside the container, the mount point locates in the Overlay2 filesystem.
\end{itemize}
The Auditd trace has the following three structures, corresponding to capability rule, network access rule and mount rule, respectively:\\

\textlangle{}apparmor\textrangle{}  \textlangle{}operation\textrangle{}  \textlangle{}profile\textrangle{} \textlangle{}pid\textrangle{}  \textlangle{}comm\textrangle{}  \textlangle{}capability\textrangle{}   \textlangle{}capname\textrangle{} \\

\textlangle{}apparmor\textrangle{}  \textlangle{}operation\textrangle{}  \textlangle{}profile\textrangle{} \textlangle{}pid\textrangle{}  \textlangle{}comm\textrangle{}  \textlangle{}family\textrangle{}   \textlangle{}sock\_type\textrangle{} \textlangle{}protocol\textrangle{} \textlangle{}requested\_mask\textrangle{} \textlangle{}addr\textrangle{} \\

\textlangle{}apparmor\textrangle{}  \textlangle{}operation\textrangle{}  \textlangle{}profile\textrangle{} \textlangle{}name\textrangle{} \textlangle{}pid\textrangle{}  \textlangle{}comm\textrangle{}  \textlangle{}fstype\textrangle{}   \textlangle{}srcname\textrangle{} \textlangle{}flags\textrangle{} \textlangle{}options\textrangle{}\\

\begin{itemize}
    \item \textbf{apparmor:} the status of AppArmor. The value for this field could be 'DENY', 'ALLOW', 'STATUS' and 'AUDIT' depending on AppArmor's status. 'AUDIT' represents that this log could be used for generating rules.  
    \item \textbf{operation:} the performed operation, which is used for distinguishing logs related to capabilities (operation\\='capable') from the ones related to network accesses.
    \item \textbf{profile:} the enforced AppArmor profile reporting the privileged attempts. Lic-Sec uses this field to find out logs for the specified container which is confined by this profile.
    \item \textbf{capname:} the name for the capability, which is added to the capability rules in the profile.
    \item \textbf{family:} supported domains for the network such as inet, inet6 and unix, which is added to the network rules in the profile.
    \item \textbf{sock\_type:}supported types for the network such as stream, dgram, and raw, which is added to the network rules in the profile.
    \item \textbf{protocol:} supported protocol numbers for the network which represent different protocols such as tcp, udp and icmp. It is also added to the network rules in the profile.
    \item \textbf{name:} the target mountpoint fileglob, should be a path, which is used for generating mount rules. 
    \item \textbf{srcname:} the source fileglob, which is used for generating mount rules.
    \item \textbf{fstype:} the allowed filesystem type such as 'ext4, debugfs, devfs', which is used for generating mount rules.
    \item \textbf{flags:} the allowed flags, such as 'ro, rw, etc', which is used for generating mount rules.
\end{itemize}

2) profile generation: the profiles are generated automatically through analyzing the trace files. Our rules generator engine grabs the key fields from the traces and combines them based on different syntax of different rules.  
\begin{itemize}
    \item \textbf{trace file:} The algorithm for translating the SystemTap trace to rules is shown as Algorithm 1. This is done by a Python script based on LiCShield’s implementation with our improvements. First, raw data of the trace file is collected line by line. For each line, we determine the applied rule (pivot root rule, link rule, file access rule, mount rule or execution rule) according to the field 'probe point name'. Second, the rules are classified into different profiles according to the fields 'control group path', 'executable process name' and 'executable path'. Third, the fields 'executable process name', 'executable path', 'resource path' and 'mount namespace root' are input to the engine to generate rules. Lic-Sec introduces a new function: when generating the execution rules for the container, it checks if the execution is for prompting shells. If no such executions are traced, it adds the deny rules for all shell commands in the container profile. Forth, the container rules are written into the AppArmor profile which has been enforced upon container’s launch instead of a newly created profile, which gives real-time container protection enforcement. 
    \item \textbf{auditd log:} The translation from the Auditd trace to capability rules, network access rules and mount rules are more straightforward using a bash script. For the capability rules and network access rules, we follow Docker-sec’s approach directly. Based on Auditd trace structures described in section 7.2.1, first, we sort out capability traces generated by the running container by searching traces with 'profile=\textlangle{}profile\_name\textrangle{}', 'apparmor=AUDIT' and 'operation=capable'. Here, the profile name is a known name of the enforced AppArmor profile for the container. Second, we get the capability used inside container from the field ’capname’ of these traces. We add them to container profile as the capability rules using the following syntax:\\
    
    capability \textlangle{}capname\textrangle{}\\
    
    In a similar way, to generate the network accesses rules, first we search out traces with 'profile=\textlangle{}profile\_name\textrangle{}' and 'apparmor=AUDIT', including the fields: 'family', 'sock type' and 'protocol'. Second, we combine the values in these three fields to generate the network access rules using the following syntax:\\
    
    network \textlangle{}family\textrangle{}   \textlangle{}sock\_type\textrangle{} \textlangle{}protocol\textrangle{}\\
    
    To generate the mount rules, first, we grab the mount operation performed by the container through searching traces with 'profile=\textlangle{}profile\_name\textrangle{}', 'apparmor=AUDIT' and 'operation=mount'. Second, we get the values in these key fields 'fstype', 'srcname', 'name' and 'flags' from those traces and combine them to a mount rule using the following syntax:\\
    
    mount fstype=\textlangle{}fstype\textrangle{} options=\textlangle{}flags\textrangle{} \textlangle{}srcname\textrangle{} -\textrangle{} \textlangle{}name\textrangle{}
\end{itemize}

\begin{algorithm}[H]
\SetAlgoLined
\KwResult{Updated container\_profile and host\_profile\ }
 input trace\_file, container\_profile, host\_profile\;
 deny\_shell\_flag = True\;
    \For{line in trace\_file}{
        applied\_rules = rules\_dispatcher(probe\_point\_name)\;
        layer = get\_layer(control\_group\_path)\;
        rules = applied\_rules(executable\_process\_name, executable\_path,  resources\_path, mount\_namespace\_root)\;
        \If{rules == 'execution\_permission\_rule' and resources\_path in ['/bin/bash', '/bin/sh', '/bin/dash'] }  {
            deny\_shell\_flag = False\;
        }
        \eIf{layer == 'container'}{
            container\_profile.add(rules)\;
        }
        {
            host\_profile.add(rules)\;
        }
    }
    \If{deny\_shell\_flag == True} {
        container\_profile.add(deny\_shell\_rules)\;
    }
    generate container profile\;
    generate host profile\;
    
\caption{Update container\_profile and host\_profile based on trace\_file}
\end{algorithm}

\subsection{Using Lic-Sec}
Lic-Sec is user-friendly to use. We provide docker-licsec commands as well as a bash script to run Lic-Sec. Lic-sec can be run in three different modes depending on the usage scenario as displayed in figure 3. Each mode has a different tracing period. Independent of the mode used, the tracing period can be terminated with the same command at any time after the container’s launch. Below, we describe the three different modes in more details:
\begin{itemize}
    \item \textbf{mode 1:} Mode 1 is designed for the scenario where comprehensive profiles for both the host and the container are needed. This mode starts SystemTap tracing before the Docker daemon is launched and starts the Auditd auditing after container is launched. Therefore, users can get all the kernel operations related to the Docker daemon, the containerd, the containerd-shim and the RunC from the very beginning of the container's life-cycle. The users need to run the bash script to enter this mode and use \textsl{docker-licsec trace-stop} to quit. 
    \item \textbf{mode 2:} Mode 2 is designed for the scenario where users only need to know the kernel operations done by the container from its initialization. This mode starts the SystemTap tracing right before the container launching and starts the Auditd auditing immediately once the container finishes initialization. Therefore, it traces the container build process by tracing the executions inside the Dockerfile specified by the image developer. Users can use the \textsl{docker-licsec run} command to enter this mode and use \textsl{docker-licsec trace-stop} to quit.
    \item \textbf{mode 3:} Mode 3 is designed for the scenario where flexible trace of the container is needed which allows users to start the tracing at any time during container's runtime. This mode is similar to the training period of Docker-sec. It only traces the container's running process. User specifies a tracing period for a specific container by running command \textsl{docker-licsec trace-start} and \textsl{docker-licsec trace-stop}, during which Lic-Sec collects information about the properties of the container. Users are supposed to perform all required executions and use all required application functionalities in order to feed necessary data to Lic-Sec and generate a profile with accurate restrictions. The tracing process is repeatable until users capture all the required functionality of the container. 
\end{itemize}

\begin{figure}[h]
\centering
    \includegraphics[width=0.65\textwidth]{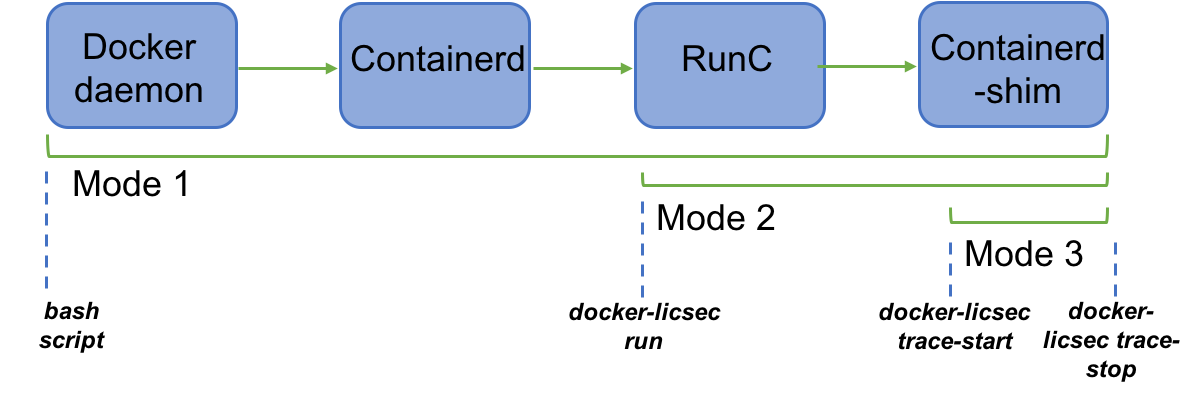}
    \caption{Lic-Sec Modes and Usage}
    \label{fig:using}
\end{figure}

\section{Security and performance evaluations of Lic-Sec}
Next, we investigate the security strength of Lic-Sec. We use the same exploit based evaluation principles as we use for the Docker-sec testing (see Section 5). We also use the very same exploit database and the main focus of the evaluation is the most severe exploits and configurations which Docker-sec did not manage to defend, i.e. the privilege escalation attacks. 
\subsection{Evaluation Result Analysis}

In the evaluation with Docker-sec, we conclude that if the capabilities and networks exploited by the attacks are truly required during container runtime, then Docker-sec cannot discard these rules hence fails to defend those kernel-targeted exploits. In the evaluation of Lic-Sec, we launched the exploits from the exploit dataset again in the Docker containers protected by Lic-Sec. We found that, compared to Docker-sec, Lic-Sec successfully defended 8 more kernel-targeted exploits aiming to gain privilege inside containers and on the host, in the Docker container running image 'Docker in Docker' and 'IBM DB2'. For other userspace program targeted attacks, Lic-Sec has the same performance as Docker-sec (see also Table \ref{tab:result1}).
 
We list the defense principles of Lic-Sec for both the privilege escalation attacks inside containers and the corresponding container escape attacks in Table \ref{tab:result2}. The EDB-ID is the ID of the privilege escalation exploits launched inside container. For the container escape attacks, as explained in Section 5.2, since we used a well proven technology and modified the original exploit, there is no official EDB-ID for those attacks. 

We have made a detailed analysis of the profile generation and protection mechanism with respect to the exploits mitigation: Since Lic-Sec does not identify any prompted shells during the tracing, it generates the corresponding deny rules for shell prompting therefore blocking the call of functions such as system() and execl(), which need to execute shell commands. The system() function uses fork() to create a child process that executes the shell command using execl('/bin/sh', 'sh', '-c', command, (char *) NULL). Since the execution of '/bin/sh' is denied by Lic-Sec, the command cannot be executed any more. For example, exploit 43418 calls system(/sbin/ifconfig lo mtu 1500) to change MTU of interface 'lo' to 1500. Exploit 41994 and 41458 call system('/sbin/ifconfig lo up') to activate interface 'lo'. Similarly, exploit 41995 calls execl('/bin/bash', 'bash', NULL) after gaining root privilege to open a shell with new Namespaces, which is also be defended by Lic-Sec. The container escape attacks are based on the corresponding gain privilege attacks inside containers, so they have the same attack principles. Therefore, due to the failed call of those functions, these exploits could not be launched as well.
\begin{table}[h]
\centering
    \caption{Lic-Sec Security Evaluation Result Overview }
    \label{tab:result2}
    \resizebox{\textwidth}{!}{
    \begin{tabular}{|c|c|c|c|c|c|c|}
    \hline
        image & \makecell{exploit\\category} & \makecell{effective\\range} & EDB-ID & CVE-ID & \makecell{vulnerable\\kernel} & Lic-Sec Defense principle \\ \hhline{|=|=|=|=|=|=|=|}
        \multirow{8}{*}{\makecell{Docker\\in\\Docker\\ \\IBM Db2}} & \multirow{8}{*}{\makecell{Gain\\privilege}} & inside container & 43418 & \multirow{2}{*} {CVE-2017-1000112} & \multirow{2}{*} {up to 4.13.9} & \multirow{6}{*} {\makecell{AppArmor denies the call of function system()\\during exploiting.}} \\ \cline{3-4}
         & & container escape & - &  &  &  \\ \cline{3-6}
         & & inside container & 41994 & \multirow{2}{*} {CVE-2017-7308} & \multirow{2}{*} {up to 4.10.6} &  \\ \cline{3-4}
         & & container escape & - &  &  &  \\ \cline{3-6}
         & & inside container & 41458 & \multirow{2}{*} {CVE-2017-6074} & \multirow{2}{*} {up to 4.9.11} &  \\ \cline{3-4}
         & & container escape & - &  &  &  \\ \cline{3-7}
         & & inside container & 41995 & \multirow{2}{*} {CVE-2016-9793} & \multirow{2}{*} {up to 4.8.13} & \multirow{2}{*} {\makecell{AppArmor denies the execution of function \\execl('/bin/bash', 'bash', NULL) after gaining root privilege.}} \\ \cline{3-4}
         & & container escape & - &  &  &  \\ \hline
    \end{tabular}
    }
\end{table}

\subsection{Performance}
We evaluated the performance for the three modes of Lic-Sec from two aspects: the profile generation time and the booting time. We collected the 10 most-used images from the Docker hub and used them for our performance measurements. When evaluating the performance for mode 1, we compared Lic-Sec with the pure LiShield since mode 1 corresponds to the LicShield tracing and profile generation principle. For mode 2 and mode 3, we compare Lic-Sec with the pure Docker-sec since these two modes are designed based on Docker-sec's framework. We ran each image with Lic-Sec and LiCShield or Docker-sec respectively for 10 times and calculated the average time after 10 rounds test for each image. 

\begin{figure}[h]
\centering
    \includegraphics[width=0.65\textwidth]{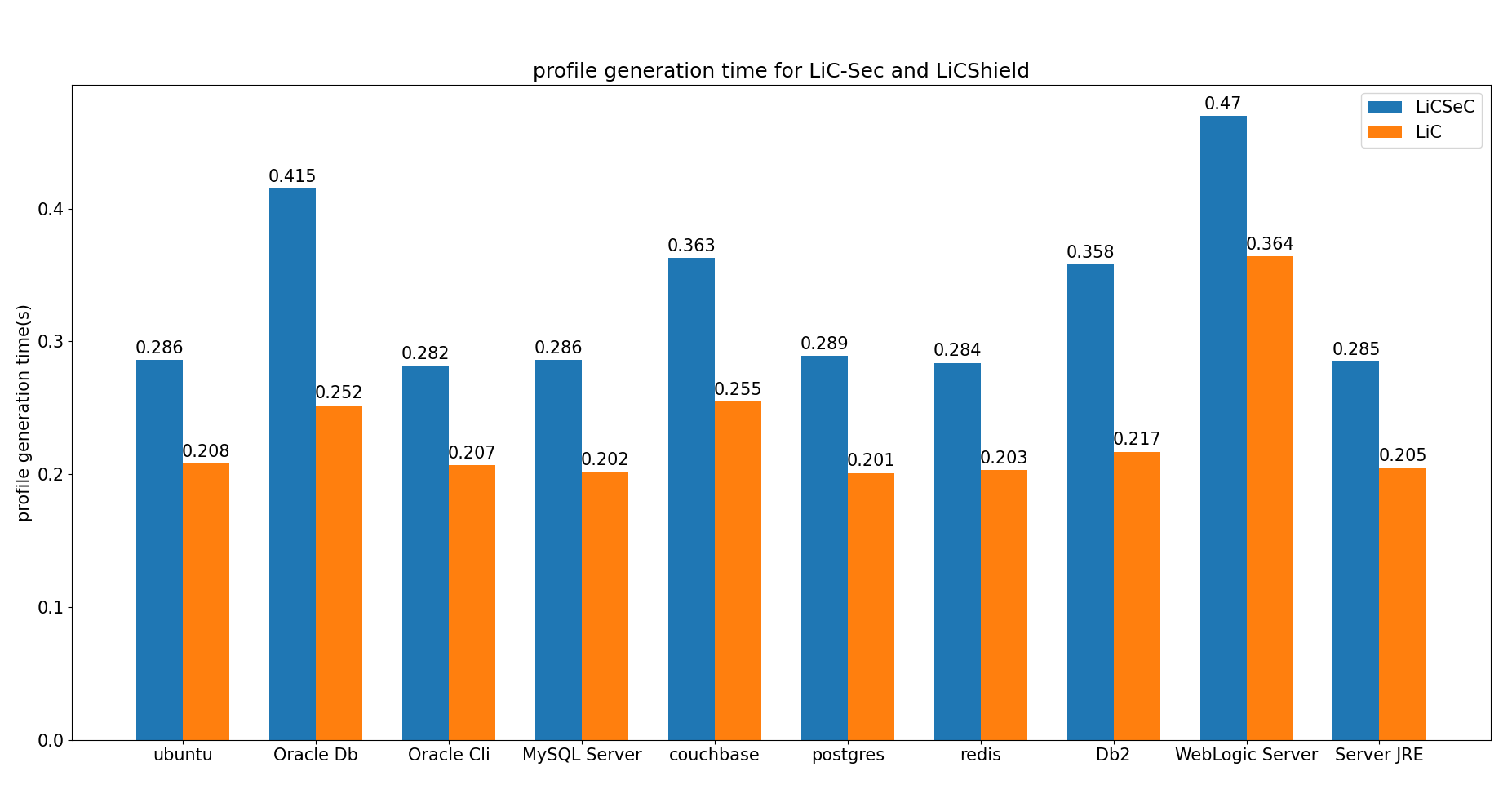}
    \caption{profile generation time for Lic-Sec and LiCShield}
    \label{fig:profile_gen}
\end{figure}

\begin{figure}[h]
\centering
    \includegraphics[width=0.65\textwidth]{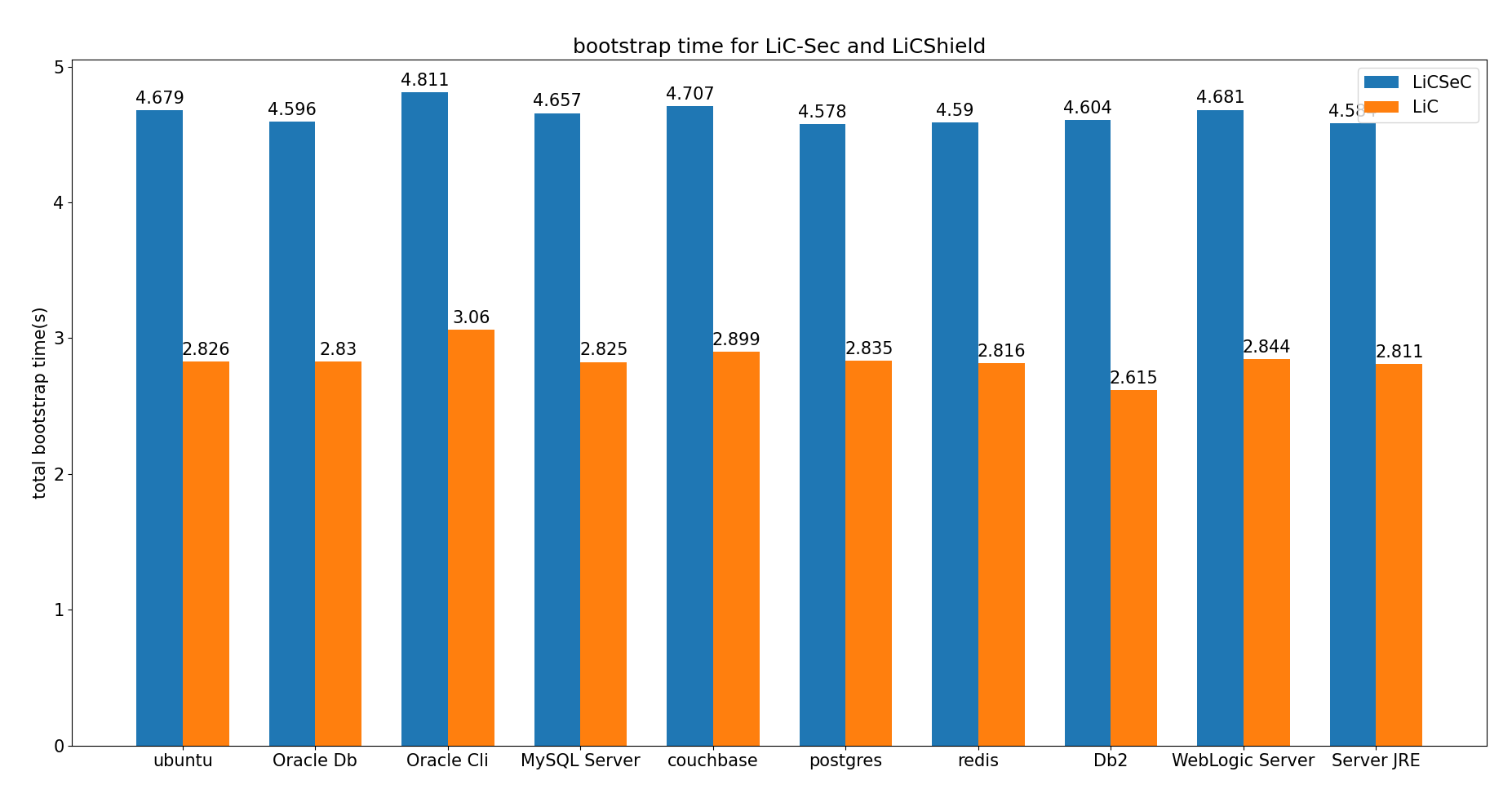}
    \caption{booting time for Lic-Sec and LiCShield}
    \label{fig:boot}
\end{figure}

\begin{figure}[h]
\centering
    \includegraphics[width=0.65\textwidth]{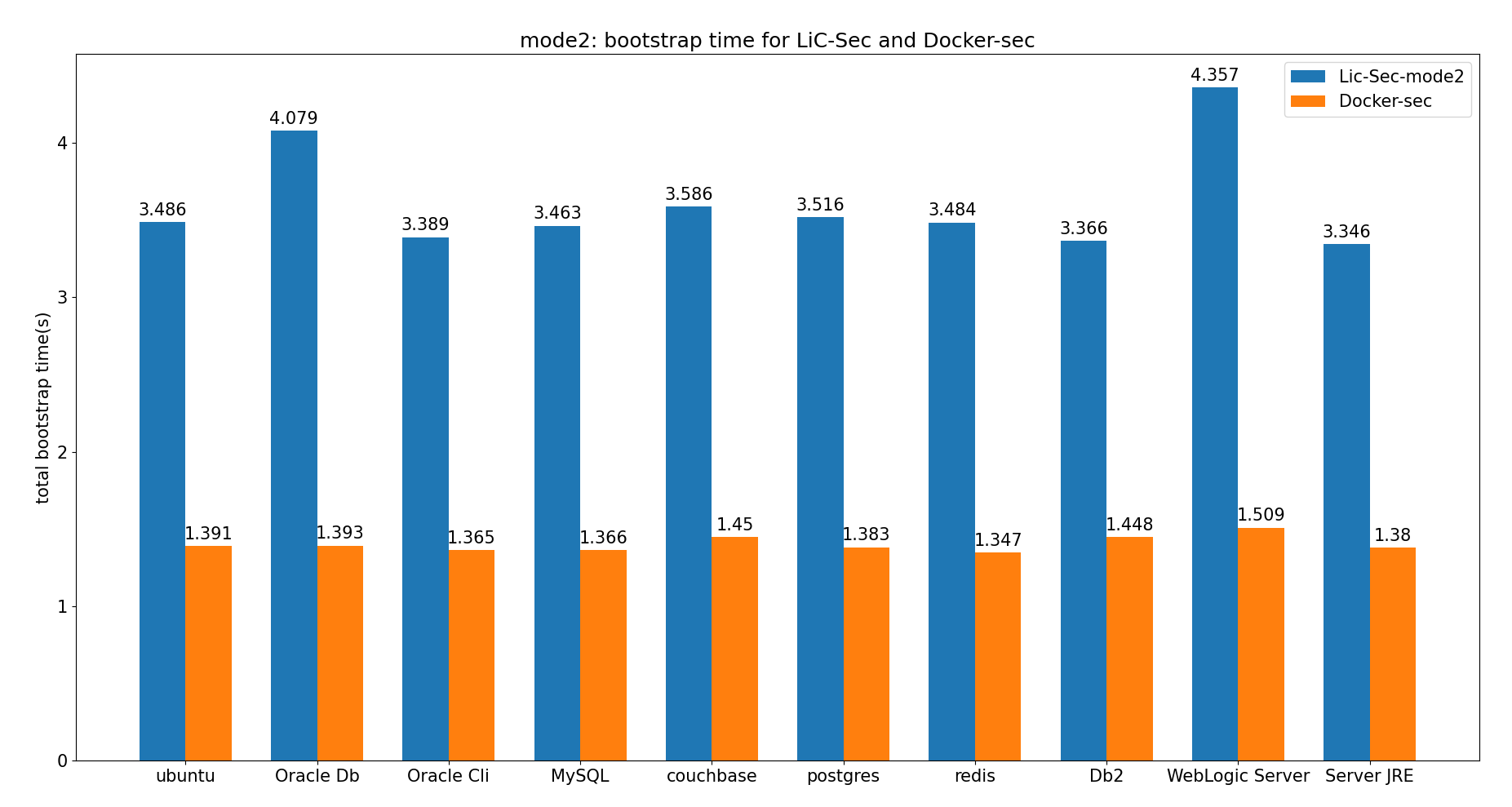}
    \caption{trace start time for Lic-Sec-mode2 and Docker-sec}
    \label{fig:run_mode2}
\end{figure}

\begin{figure}[h]
\centering
    \includegraphics[width=0.65\textwidth]{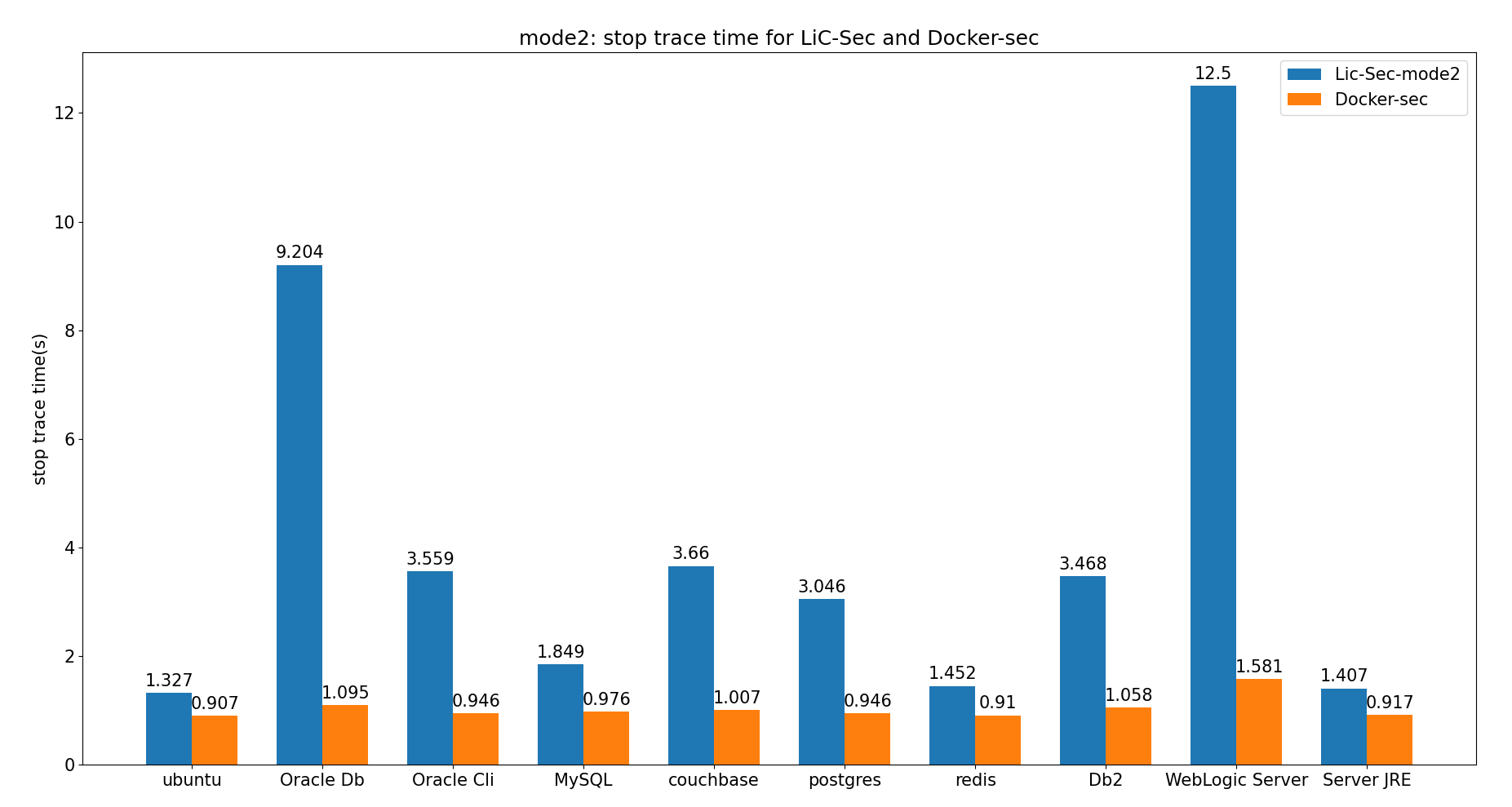}
    \caption{trace termination time for Lic-Sec-mode2 and Docker-sec}
    \label{fig:stop_mode2}
\end{figure}

\begin{figure}[h]
\centering
    \includegraphics[width=0.65\textwidth]{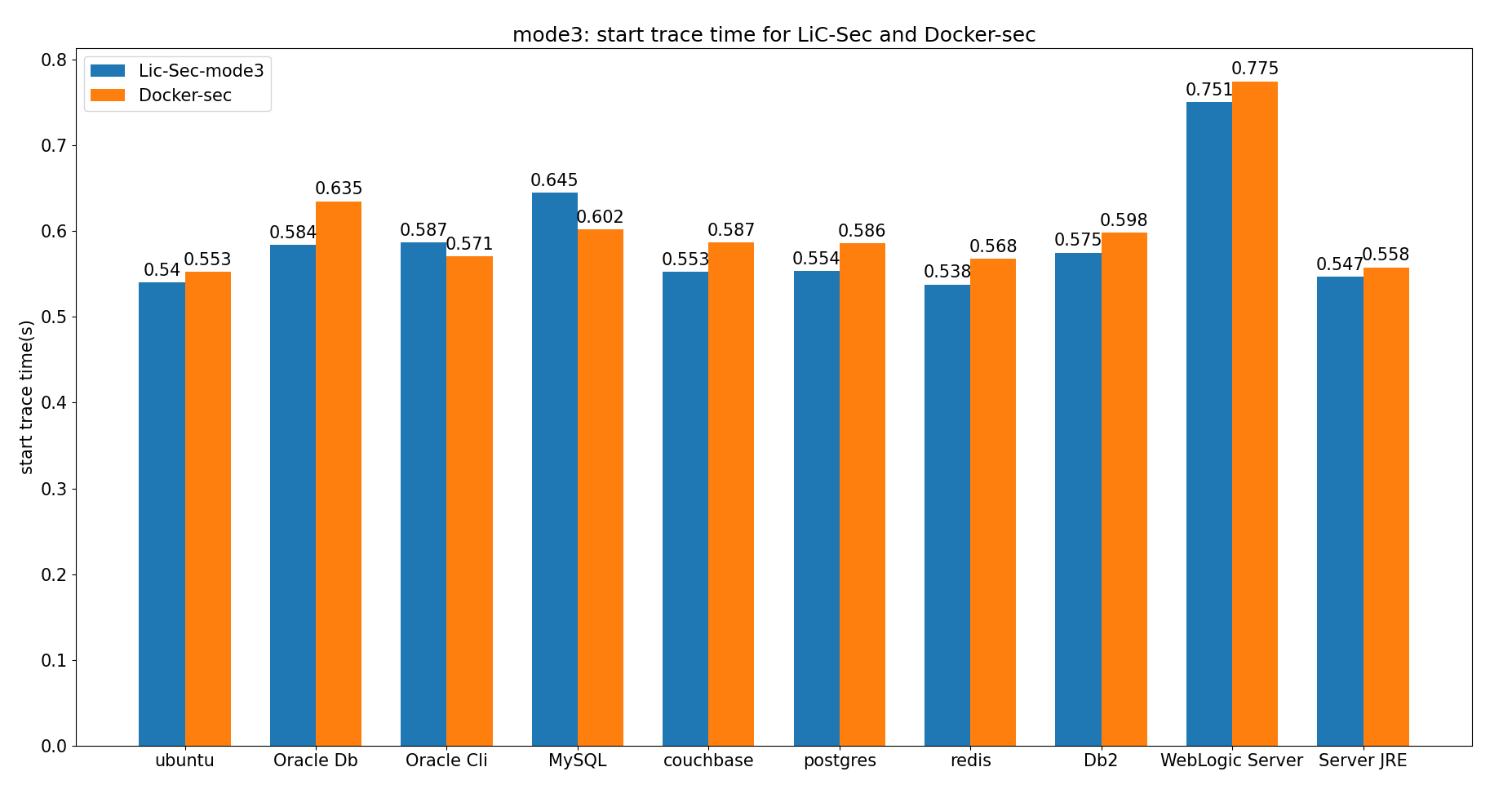}
    \caption{trace start time for Lic-Sec-mode3 and Docker-sec}
    \label{fig:start_mode3}
\end{figure}

\begin{figure}[h]
\centering
    \includegraphics[width=0.65\textwidth]{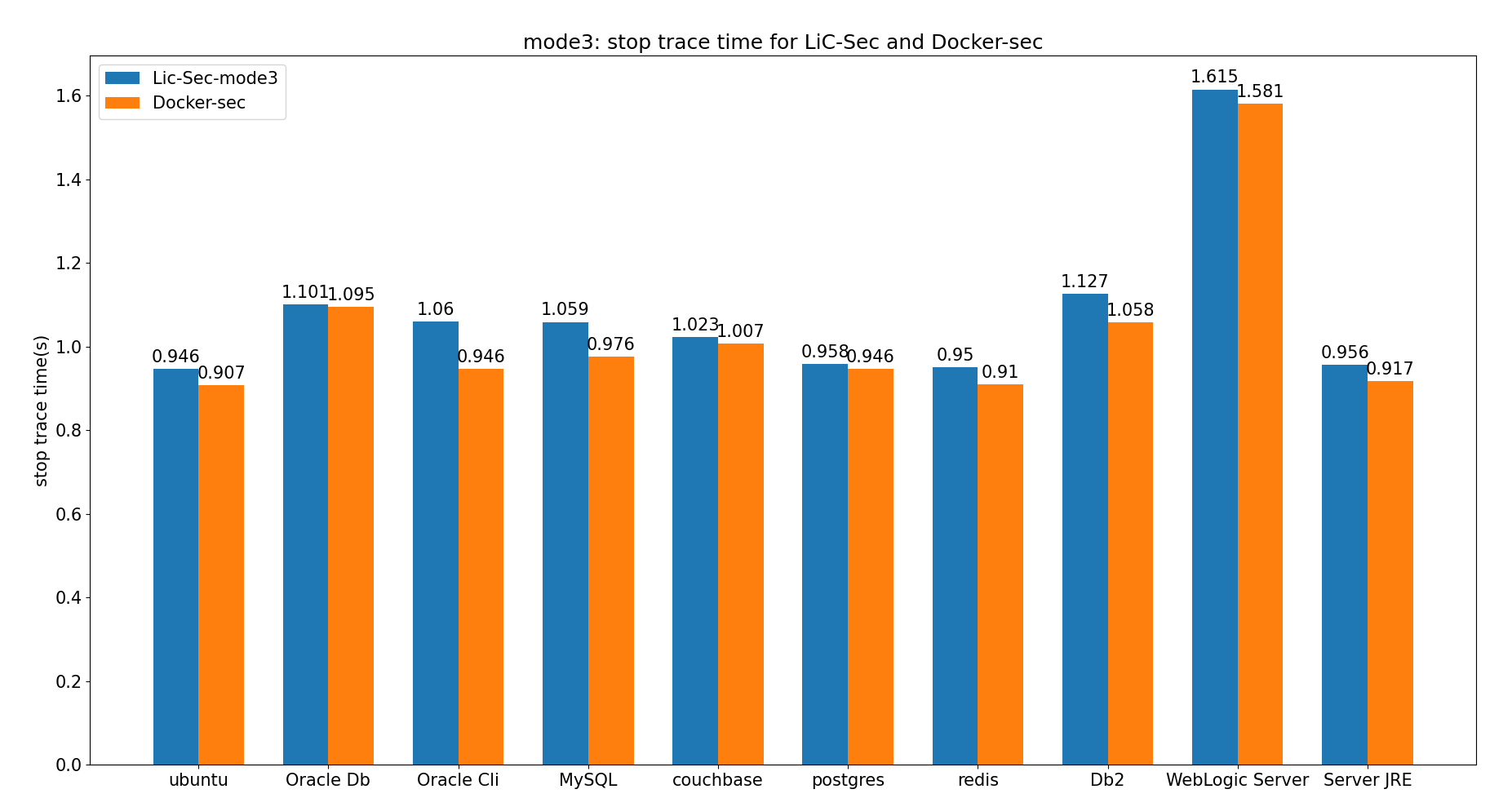}
    \caption{trace termination time for Lic-Sec-mode3 and Docker-sec}
    \label{fig:stop_mode3}
\end{figure}

The results for Lic-Sec mode1 and pure LiCShield are shown in figure \ref{fig:profile_gen} and figure \ref{fig:boot}. The overhead introduced by Lic-Sec regarding profile generation time is nearly ignorable. In terms of the bootstrap time, Lic-Sec introduces much more overhead due to the running of Docker-sec in parallel. The comparison of execution time for starting and terminating the trace between Lic-Sec's mode2 and Docker-sec are shown in figure \ref{fig:run_mode2} and figure \ref{fig:stop_mode2}. Lic-Sec utilizes 2 more seconds in average then Docker-sec to start the trace, which is caused by the running of SystemTap trace. When SystemTap is set up, it first compiles the trace script into kernel modules and loads them into the kernel, which introduces some extra time. Considering the time for terminating the trace, Lic-Sec mode2 consumed more time than Docker-sec in general but showed different performances depending on images. For images such as ubuntu, redis and server JRE, Lic-Sec mode2 didn't introduce obvious extra execution time. However, for images such as weblogic server and Oracle DB, Lic-Sec mode2 consumed much more time to stop the trace. This increase in time is due to the fact that many more kernel operations were traced by Lic-Sec during the launch time for these images, which increased the workload of profile generator engine. The comparison of execution time for starting and terminating the trace between Lic-Sec's mode3 and Docker-sec are shown in figure \ref{fig:start_mode3} and figure \ref{fig:stop_mode3}. Considering the trace termination time in this case, we found that Lic-Sec performed basically the same as Docker-sec. The reason is that, Lic-Sec mode3's tracing starts after the container has been launched so kernel operations performed for launching the container are not traced. Since a large proportion of the trace is recorded during the launch time of the container, workload is not increased significantly to the profile generator engine if trace starts during container's runtime, which introduces nearly ignorable overhead to the trace termination time.

\section{Related Work}
\subsection{Container security solutions}
Several previous research work address LSMs for Docker containers. Some work focus on enhancing security by applying customized LSM modules’ policies such as AppArmor and SELinux. Docker-sec \cite{loukidis2018docker} and LiCShield\cite{mattetti2015securing} which are discussed in this paper are two of them. Bacis et al. proposed a solution based on SELinux hardening with an SELinux policy module added to the Dockerfile. In this way, containerized processes are run with SELinux types by binding SELinux policies with Docker container images, which specializes SELinux policy per-container or even per-process to increase Docker security \cite{bacis2015dockerpolicymodules}. This kind of approach relies on the host to define and enforce policies which infers that only the system administrator can use these approaches.
Other than tailoring AppArmor and SELinux policies for each container to enhance security, researchers are exploring other ways to enable containers to use LSM modules to improve their own security. Y. Sun et al. proposed the design of security namespace, which is a kernel abstraction that enables containers to utilize a virtualization of the Linux kernel security framework to achieve autonomous per-container security control \cite{sun2018security}. The main difference compared to our work is that, first, security namespace 
enables the container to autonomously control its security rather than relying on the system administrator to enforce the security control from the host. In our work, the policies are defined and enforced solely by the host. In \cite{sun2018security}, policies are pushed as string rules from the container directly to the kernel without the involvement of the host. Second, the whole kernel security framework could be used by containers rather than just limited security features since security namespace virtualizes kernel security frameworks into virtual instances, each of which corresponds to one container. However, security namespace requires intensive modifications of the kernel base and the LSM modules, and the policies are not automatically generated. The experimental results indicated that the security Namespaces can solve several container security problems with acceptable performance overhead. DIVE (Docker Integrity Verification Engine) \cite{de2019integrity} is an architecture to support integrity verification and remote attestation of Docker containers. DIVE relies on a modified version of IMA (Integrity Measurement Architecture) which is also a widely used LSM module \cite{sailer2004design}, and OpenAttestation which is a well-known tool for attestation of cloud services. With the help of DIVE, the infrastructure manager can be informed of the specific compromised container or hosting system and request to rebuild this single container. DIVE is practical to deploy due to a nearly negligible performance impact on the hosted services. 
Besides using LSM modules to improve Docker container security, kernel-based solutions are also proposed \cite{priedhorsky2017charliecloud} \cite{azab2017enabling}. These solutions aim to provide a secure framework or wrapper to run Docker container. In the work of Reid and Tim \cite{priedhorsky2017charliecloud}, a secure framework based on the Linux user and mount namespaces called Charliecloud is proposed to run industry-standard Docker containers without privileged operations or daemons on center resources. Charliecloud is proved to avoid most security risks such as bypass of file and directory permissions and chroot escape. In the work of \cite{azab2017enabling}, a secure wrapper called Socker is described for running Docker containers on Slurm and similar queuing system. The execution of containers within Slurm jobs is enforced by Socker as the submitting user instead of root user to avoid the privileged operations. Socker also bounds the resource usage of any container by the amount of resources assigned by Slurm.

There are some researches proposing security countermeasure or algorithm against a particular attack category of container. This includes special investigations on DoS attacks \cite{chelladhurai2016securing}, container escape attacks \cite{jian2017defense}, attacks from the underlying compromised higher-privileged system software such as the OS kernel and the hypervisor \cite{arnautov2016scone}, covert channels attacks \cite{luo2016whispers} and the application level attacks \cite{hunger2018dats}. In \cite{chelladhurai2016securing}, Chelladhurai et. al apply three-tier protection mechanism including memory limit assignment, memory reservation assignment and default memory value setting to defend against DoS attacks. The main principle behind their mechanism is to limit the resource consumption of the container. In \cite{jian2017defense}, similar to our work, Jian et al. also investigated Docker escape attacks but particularly focusing on Namespace switching escapes. They proposed a defense based on Namespaces status inspection 
against this kind of attacks. According to their solution, once a different Namespaces tag is detected, the corresponding process will be killed immediately and the malicious user will be tracked. The test results showed that this defense can effectively prevent some real-world attacks. In \cite{arnautov2016scone}, SCONE is proposed, which is a secure container environment for Docker utilizing Intel Software Guard eXtension (SGX) \cite{hoekstra2013using} to run Linux applications in secure containers. The solutions provides confidentiality and integrity of application data within containers. Covert channels attacks against Docker containers are analyzed in \cite{luo2016whispers}. In this paper, Luo et al. identify different types of covert channel attacks in Docker and propose solutions of using and configuring current security mechanisms provided by Docker to prevent those attacks. They also stress that the deployment of a full-fledged SELinux or AppArmor security policy is a key condition to protect the security perimeters of containers. Study about increase the application level isolation for containers is presented in the work of Hunger et al. \cite{hunger2018dats}. The authors propose DATS which is a system to run web applications that heavily access data in shared folders and that are deployed with containers. The system enforces non-interference across containers and is effective to mitigate all data-disclosure vulnerabilities.

Some researches address network security challenges for Docker containers. In the work of Ranjbar et. al \cite{ranjbar2017synaptic}, they propose SynAPTIC architecture to  provide  secure  and  persistent  connectivity between containers based on the standard host identity protocol (HIP). Cilium \footnote{https://github.com/cilium/cilium} is an open source software for transparently securing the network connectivity between application services deployed using Linux container management platforms such as Docker and Kubernetes. In the work of Secure Cloud \cite{kelbert2017securecloud}, secure communication among containerized services is realized with the support of Intel’s Software Guard Extensions (SGX).

\subsection{Container security testing and evaluation}
Besides proposing security solutions and enhancements, previous researches also present, similar to our work, evaluation of container security. Most of them evaluate container security theoretically from the perspective of system architecture and design principle \cite{babar2017understanding} \cite{casalicchio2020state} \cite{yasrab2018mitigating} \cite{mp2016enhancing} \cite{manu2016study}\cite{sultan2019container}. In \cite{lin2018measurement},  Lin et. al, described the first measurement study of Docker container's security. We have used a similar methodology in our investigation, but tailored it to the Docker-sec evaluation target.

Some related researches are about early detection of Docker vulnerabilities such as performing penetration tests \cite{lu2017research}, implementing static code analyzer \cite{duarte2018empirical}, implementing static and dynamic vulnerability detection schemes \cite{tunde2019study} and analyzing docker image vulnerabilities \cite{shu2017study}\cite{zerouali2019impact}\cite{brady2020docker}. In \cite{lu2017research}, Tao et. al clarified the importance of penetration testing technology in ensuring the container security and studied some typical penetration test cases such as container escape and DoS in container environment. Then they described the penetration testing process in Docker container in details. In \cite{duarte2018empirical}, Ana et. al analyzed the applicability of static code analyzers in Docker code base and concluded that these tools are very ineffective to detect Docker vulnerabilities. But they got the same conclusion as in \cite{lu2017research}, i.e. Docker vulnerabilities, especially bypass and privilege escalations, are easy to find using penetration testing. A study on the effectiveness of various vulnerability detection schemes for containers is conducted in \cite{tunde2019study} using 28 real world vulnerability exploits. Their results showed that dynamic detection scheme performs much better than static detection scheme. Combining static and dynamic schemes can further increase the detection coverage. The authors in \cite{shu2017study} analyze the Docker Hub images by using the framework DIVA (Docker Image Vulnerability Analysis). They studied 356,218 images and found that both official and community images contain more than 180 vulnerabilities on average. They recommend that automatic security updates could solve this problem. In \cite{zerouali2019impact}, an empirical study about the impact of npm JavaScript package vulnerabilities in Docker images was conducted. The results showed that the outdated npm packages in Docker images pose a risk of security vulnerabilities so that the installed JavaScript packages should be up to date to avoid potential security risks from Docker images. In \cite{brady2020docker}, Brady et. al, described a continuous integration and continuous deployment (CI/CD) system to evaluate the security of Docker images under the scenario when container is used as part of the software development. The results showed that the system is effective to prevent publishing and reuse of images with known vulnerabilities.

\section{Conclusion and Future Work}
In this paper, we evaluated the security defense strength of Docker-sec by testing real-world attacks on containers. We manually collected 42 exploits out of 400 exploits from Exploit-db. The selected exploits were all effective on container platforms with the default Docker security configurations. We categorized the exploits into two categories,  kernel-targeted exploits and user space program targeted exploits. 

Using the selected exploits,  we tested the top 20 official images from Docker registry hub in containers spawned with Docker-sec. We ran the test on a host with a vulnerable kernel. Our investigation shows that for 18 out of 20 images, Docker-sec can defend against all attacks in the kernel target category, which aim at escalating privileges inside or outside the container. Only for two of the tested images, 'Docker in Docker' and 'IBM DB2', Docker-sec was almost chance-less. This result clearly shows the benefits and limitations with Docker-sec. The main strength with the approach is the capability to release the end-user burden in configuring strict container security polices as this is instead automated by Docker-sec. The resulting strict LSM gives a reasonable level of security against kernel vulnerabilities. Still, as clearly shown in our evaluation, for applications running on the container which require a broad set of privileges, the Docker-sec approach does not give much additional security value. 

We also tested vulnerable user space programs inside containers spawned with Docker-sec and showed that Docker-sec failed to defend the attacks exploiting vulnerabilities of a specific user space program. This is mainly due to the fact that it can be launched independent of the Linux capabilities and of network accesses. However, it is worth noticing, that even if Docker-sec was not directly effective against these types of vulnerabilities, the default read-only restriction for file access still blocks propagation to the host from the compromised container.

To address the issues which couldn't be solved by Docker-sec, we introduced an earlier proposed LSM called LiCShield due to its strengths which Docker-sec lacks, i.e, LiCShield generates AppArmor profiles automatically based on the traced kernel operations hence generated more types of rules. We combined Docker-sec with a modified version of LiCShield to construct a new AppArmor profile generator called Lic-Sec. Besides auditing capabilities and network accesses, Lic-Sec was able to generate pivot\_root\_rules, link\_rules, file\_access\_rules, mount\_rules and execution\_rules. We evaluated the effectiveness of Lic-Sec against real-world exploits in the same way and made a comparative analysis of the evaluation results and performance of Docker-sec and Lic-Sec. The results showed that Lic-Sec was effective against defending privilege escalation attacks and could block the execution and propagation of all tested attacks with reasonable performance overhead. Although the security evaluation of Lic-Sec is performed based on an old kernel in order to make all kernel exploits available, we consider the experimental evaluation results of Lic-Sec of protecting old kernels against privilege escalation attacks could indicate that Lic-Sec has the potential to defend against zero-day attacks in the future.

It is left to future work to compare Lic-Sec against other LSM based protection approaches such as the ones presented in \cite{sun2018security}\cite{bacis2015dockerpolicymodules}. Then we will get an even better understanding of the strength of Lic-Sec compared with alternative available tools. Also, in order further assess the strength/weaknesses of  Lic-Sec, it remains to test it using penetration testing tools and additional exploit datasets.

%
%
%
\bibliographystyle{splncs04}
\bibliography{licsec.bib}

\end{document}